\documentclass[preprint]{imsart}
\RequirePackage[OT1]{fontenc}
\RequirePackage{amsthm,amsmath}
\RequirePackage[numbers]{natbib}
\RequirePackage[colorlinks,citecolor=blue,urlcolor=blue]{hyperref}
\usepackage{amssymb, amsmath,  amsthm, graphicx, euscript, mathrsfs, enumerate}
\def\I{I\!\!I}
\renewcommand{\i}{\mathrm{i}}

\newcommand{\eqd}{\stackrel{d}{=}}

\newcommand{\cov}{\operatorname{cov}}

\newcommand{\E}{{\mathbb E}}

\renewcommand{\L}{{\EuScript L}}

\newcommand{\Var}{\operatorname{Var}}
\renewcommand{\Re}{\mathrm{Re}}
\renewcommand{\Im}{\mathrm{Im}}

\newcommand{\M}{\mathcal{M}}
\newcommand{\NN}{\mathcal{N}}
\theoremstyle{plain}
\newtheorem{thm}{Theorem}[section]

\newtheorem{cor}[thm]{Corollary}

\theoremstyle{definition}

\newcommand{\C}{{\mathbb C}}
\newcommand{\R}{{\mathbb R}}

\newcommand{\hmu}{\widehat{\mu}}
\newcommand{\hg}{\widehat{g}}
\newcommand{\hpsi}{\widehat{\psi}}

\begin{document}
\pubyear{\today}
\begin{frontmatter}
\title{Semiparametric estimation   in the normal variance-mean mixture model\thanksref{T1}}
\runtitle{Estimation in the variance-mean mixture model}

\begin{aug}
\author{\fnms{Denis} \snm{Belomestny}\ead[label=e1]{denis.belomestny@uni-due.de}}
\address{University of Duisburg-Essen\\
Thea-Leymann-Str. 9, 45127 Essen,  Germany\\
and \\
Laboratory of Stochastic Analysis and its Applications \\
             National Research University Higher School of Economics\\
             Shabolovka, 26, 119049 Moscow,   Russia\\
\printead{e1}}

\author{\fnms{Vladimir} \snm{Panov}\ead[label=e2]{vpanov@hse.ru}}
\address{Laboratory of Stochastic Analysis and its Applications \\
             National Research University Higher School of Economics\\
             Shabolovka, 26, 119049 Moscow,   Russia\\
\printead{e2}}
\thankstext{T1}{
This work has been funded by the  Russian Academic Excellence Project ``5-100''. 
}
\runauthor{D.Belomestny and V.Panov}
\end{aug}

\begin{abstract}
In this paper we study the problem of statistical inference on the parameters of the semiparametric variance-mean mixtures.  This class of mixtures has recently become rather popular in statistical and financial modelling. We design a semiparametric estimation procedure that first estimates the mean of the underlying normal distribution and then recovers nonparametrically the density of the corresponding mixing distribution.  We illustrate the performance  of our procedure on simulated and real data.
\end{abstract}
\tableofcontents
\end{frontmatter}
\section{Introduction and set-up}
A normal variance-mean mixture is defined as
\begin{eqnarray}
\label{vm}
p(x; \mu, G)=
	\int_{\R_{+}} \varphi_{\NN(\mu s, s)} (x) \, G(ds)
=
\int_{\R_{+}} \frac{1}{\sqrt{ 2 \pi s}} \exp\left\{
	- \frac{(x - \mu s)^{2}}{2s}
\right\} \, G(ds),
\end{eqnarray}
where \(\mu\in \mathbb{R},\)  \(\varphi_{\NN(\mu s, s)} \) stands for the density of a normal distribution with mean \(\mu s\) and variance \(s\),  and  $G$ is a mixing distribution on $\mathbb{R}_{+}.$   As can be easily seen, a random variable \(X\) has the distribution \eqref{vm} if and only if
\begin{eqnarray}
\label{X}
X\eqd\mu \xi + \sqrt{\xi} \eta,\quad \eta\sim \mathcal{N}(0,1),\quad \xi \sim G.
\end{eqnarray} 
The variance-mean
mixture models play an important role in
statistical modelling and have many applications. In particular, such mixtures appear as limit distributions in
the asymptotic theory for dependent random variables and they are also useful for
modelling data stemming from heavy-tailed and skewed distributions, see, e.g. Barndorff-Nielsen, Kent and S{{\o}}rensen  \cite{BNKS}, Barndorff-Nielsen \cite{bn97}, Bingham and Kiesel \cite{BK}, Bingham, Kiesel and Schmidt \cite{BKS}.  
If \(G\) is the generalized inverse Gaussian distribution, then the normal variance-mean mixture distribution coincides with the so-called generalized hyperbolic distribution. The latter distribution has an important property that the logarithm of its density function is a smooth unimodal  curve approaching linear asymptotes. This type of distributions was used to model the sizes of the particles of sand (Bagnold \cite{Bagnold}, Barndorff-Nielsen and Christensen \cite{BNC}), or  the diamond sizes in marine deposits in South West Africa (Barndorff-Nielsen \cite{BN77}). 
\par
In this paper we study the problem of statistical inference for the mixing distribution \(G\) and the parameter \(\mu\) based on a sample \(X_1,\ldots, X_n\) from the distribution with density \(p(\cdot; \mu, G).\) This problem was already considered in the literature, but mainly in the parametric situations. For example, in the case of  the generalised hyperbolic distributions some parametric approaches can be found in J{{\o}}rgensen \cite{Jorgensen}, and  Karlis and Lillest{\"o}l \cite{kl}.
There are also  few papers dealing with the  general semiparametric case.  For example, Korsholm
\cite{Korsholm} considered the statistical inference  for a more general model of the form
\begin{eqnarray}
\label{Kors}
p_{K}(x; \mu, G)=
	\int_{\R_{+}} \varphi_{\NN(\alpha + \beta/ s  ,  1/ s)} (x) \, G(ds)
\end{eqnarray}
and proved the consistency of the non-parametric
maximum likelihood estimator for the parameters $\alpha$ and $\beta$, whereas $G$  was treated as an nuisance probability distribution.  Although the maximum likelihood (ML) approach of  Korsholm is rather general, its practical implementation would meat serious computational difficulties, since one would need to solve rather challenging optimization problem.  Note that the ML approach for similar models  was also considered by van der Vaart \cite{vaart}.
Among other papers on relevant topic, let us mention the paper by Tjetjep and Seneta \cite{ts}, where the method of moments was used for some special cases of the  model \eqref{vm}, and the paper by  Zhang \cite{Zhang}, which is devoted to  the problem of estimating the mixing density in  location (mean) mixtures.
\par
The main contribution of this paper is a new computationally efficient estimation approach which can be used to estimate both the parameter \(\mu\) and the mixing distribution \(G\) in a consistent way. This approach employs the Mellin transform technique and doesn't involve any type of high-dimensional optimisation. We show that while our estimator of \(\mu\) converges with parametric rate, a nonparametric estimator of the density of \(G\) has much slower convergence rates.

\par
The paper is organized as follows. In Section~\ref{seq:mu} the problem of statistical  inference for \(\mu\) is studied.  Section~\ref{sec4} is devoted to the estimation of $G$ under known $\mu$ and Section~\ref{sec5}  extends the results of Section~\ref{sec4} to the case of unknown \(\mu.\) A simulation study is presented in Section~\ref{num} and a real data example can be found in Section~\ref{rdex}.

\section{Estimation of $\mu$} 
\label{seq:mu}
First note that the density in the normal variance-mean model can be represented in the following form 
\begin{eqnarray}
\label{p1}
p(x; \mu, G)=
	e^{x \mu} I_{\mu,G}\left( - x^{2} / 2\right), \; 
\end{eqnarray}
where
\begin{eqnarray*}
	I_{\mu,G} (u) := 
\int_{\R_{+}} \frac{1}{\sqrt{ 2 \pi s}} \exp\left\{
	\frac{u}{s} - \frac{\mu^{2}}{2} s
\right\} \, G(ds).
\end{eqnarray*}
This observation in particularly implies that  
\begin{eqnarray}
\label{p2}
p(-x; \mu, G)=
	e^{-x \mu} I_{\mu,G}\left( - x^{2} / 2\right), \; 
\end{eqnarray}
and therefore, dividing  \eqref{p1} by \eqref{p2}, we get 
\begin{eqnarray}
\label{eq:mu_repr}
	\mu =  \frac{1}{2x} \log \left( 
	\frac{p(x; \mu, G)}{p(-x; \mu, G)}
	\right).
\end{eqnarray}
The formula \eqref{eq:mu_repr} represents \(\mu\) in terms of \(p(\cdot; \mu, G).\) 
The representation \eqref{eq:mu_repr} can also be written in the form 
\begin{eqnarray}
\label{mu}
	\mu = \int_{\R}  \frac{1}{2x}  \;\log \left( 
	\frac{p(x; \mu, G)}{p(-x; \mu, G)}
	\right) \; p(x; \mu, G)\, dx,
\end{eqnarray}
which looks similar to the entropy of \(p(\cdot; \mu, G):\)
\begin{eqnarray*}
 H(p) := -   \int_{\R}  \log \left( 
		p(x; \mu, G)
	\right) \, p(x; \mu, G) dx.
\end{eqnarray*}
For a comprehensive overview of the methods of estimating \(H(p)\), we refer to \cite{BDGM}. Note also that the estimation of the functionals like \eqref{mu} was considered in \cite{BL}. Typically, the parametric convergence rates for the  estimators of such functionals can be achieved only under very restrictive assumptions on the density   \(p(x).\)  In an approach presented below, we avoid these restrictive  conditions and prove a square root convergence under very mild assumptions.
Let $w(x)$ be a Lipschitz continuous function on $\mathbb{R}$ satisfying
\[
w(x)\leq0,\quad x\geq0,\quad w(-x)=-w(x),\quad\mathrm{supp}(w)\subset[-A,A]
\]
for some $A>0.$ Set 
\[
W(\rho):=\mathbb{E}\left[e^{-\rho X}w(X)\right],\quad\rho\in\mathbb{R},
\]
then the function $W(\rho)$ is monotone and $W(\mu)=0.$ This property
suggests the following method to estimate $\mu.$ Without loss of
generality we may assume that $\mu\in[0,M/2)$ for some $M>0.$ Set
\begin{eqnarray}
\label{wmu}
\mu_{n}:=\inf\{\rho>0:\,W_{n}(\rho)=0\}\wedge M
\end{eqnarray}
with 
\[
W_{n}(\rho):=\frac{1}{n}\sum_{i=1}^{n}e^{-\rho X_{i}}w(X_{i}).
\]
Note that since $\lim_{\rho\to-\infty}W_{n}(\rho)\leq0,$ $\lim_{\rho\to\infty}W_{n}(\rho)\geq0$
and
\[
W_{n}'(\rho)=\frac{1}{n}\sum_{i=1}^{n}(-X_{i})e^{-\rho X_{i}}w(X_{i})\geq0
\]
for all $\rho\in\mathbb{R},$ the function $W_{n}(\rho)$ is monotone
and $\mu_{n}$ is unique. The following theorem describes the convergence properties of \(\mu_n\) in terms of the norm 
\(\left\Vert \mu-\mu_{n}\right\Vert _{p} := \left(
\E 
\left|
\hat{\mu}_{n} -\mu
\right|^{p}
\right)^{1/p},\) where \(p \geq 2.\)
\begin{thm}
\label{thm:murates}
Let \(p\geq 2\) and \(M>0\) be such that $$\Lambda(M,p):=\left\Vert 
\left(
1+ e^{-MX}
\right) Xw(X)\right\Vert _{p}<\infty.$$
Then 
\begin{align*}
\left\Vert \mu_{n} - \mu\right\Vert _{p}  \leq\frac{KM^{1-1/p}}{n^{1/2}}\left[\frac{1}{W'(\mu)}+\frac{1}{W(M/2)}\right]
\end{align*}
with a constant \(K\) depending on \(p\) and \(\Lambda(M,p)\) only.
\end{thm}
\section{Estimation of $G$ with known $\mu$} 
\label{sec4}
 In this section, we assume that the distribution function \(G\) has a Lebesgue density \(g\) and our aim is to estimate \(g\) from the i.i.d. observations \(X_{1},\ldots, X_{n}\) of the random variable \(X\) with the density \(p(x; \mu, G),\) provided that the parameter \(\mu\) is known.
 The idea of the estimation procedure is based on the following observation. Due to the representation \eqref{X},  the characteristic function of \(X\) has the form:
\begin{eqnarray}\label{phix}
\phi_{X}(u) :=\E \left[ 
	e^{\i u X}
\right] =\E \left[
e^{\xi \psi(u)} 
\right]
=
\L_{\xi} \left(
\psi(u)
\right),
\end{eqnarray}
where \(\L_{\xi} (x) :=\E [ e^{-\xi x} ] = \int_{\R} e^{-sx} g(s) \,ds \) is the Laplace transform of the r.v. \(\xi\) and \(\psi(u) = -  \i  \mu u + u^{2}/2\) is the characteristic exponent of the normal r.v. with mean \(\mu\) and variance \(1.\) 
Our approach is based on the use of the Mellin transform technique. Set
\begin{eqnarray*}
\M \left[
	\L_{\xi}
\right]  (z) := 
\int_{\R_{+}}   \L_{\xi}(u) u^{z-1} du,
\end{eqnarray*}
then by the integral Cauchy theorem   
\begin{eqnarray*}
\int_{\R_{+}}   \L_{\xi}(u) u^{z-1} du = 
\int_{l}   \L_{\xi}(w) w^{z-1} dw,
\end{eqnarray*}
where \(l\) is the curve on the complex plane defined as  the image of \(\psi(u)\) by mapping from \(\R\) to \(\C,\) that is, \(l\) is the set of points \(z\in \C\) satisfying  \(\Im(z) = - \mu \sqrt{2 \; \Re(z)}.\) Therefore, we get 
\begin{eqnarray*}
\M \left[
	\L_{\xi}
\right]  (z)
 =
\int_{\R_{+}} \L_{\xi}(\psi(u))  \left[\psi(u)\right]^{z-1}  \psi'(u) du =
\int_{\R_{+}}   \phi_{X}(u) \left[\psi(u)\right]^{z-1} \psi'(u) du
\end{eqnarray*}
so that the Mellin transform \(\M \left[
	\L_{\xi}
\right](z)\) can be  estimated from data via
\begin{eqnarray}
\label{hatM}
\widehat{\M} \left[
	\L_{\xi}
\right](z)  
:=
\begin{cases}
\int_{0}^{U_{n}}  \phi_n(u)\left[ \psi(u) \right]^{z-1} \psi'(u) du, & \mu \; \Im(z) <0, \\
\int_{0}^{U_{n}}  \overline{\phi_n(u)} \left[ \overline{\psi(u)} \right]^{z-1} \overline{\psi'(u)} du, & \mu \; \Im(z)>0,
	\end{cases}
\end{eqnarray}
where 
\begin{eqnarray*}
\phi_n(u)=\frac{1}{n} \sum_{k=1}^{n} e^{\i u X_{k}}, 
\end{eqnarray*}
and \(U_{n}\) is sequence of positive numbers tending to infinity  as \(n \to \infty.\) This choice of the estimate for \({\M} \left[
	\L_{\xi}
\right](z)\) is motivated by the fact that the function
\begin{eqnarray*}
\left| 
\left[
	\psi(u)
\right]^{z}
\right| = 
	\exp\left\{
		(\Re(z)/2) \log \left(
			\mu^{2} u^{2} 
			+
			u^{4} / 4
		\right)
	\right\}
	\cdot  
	\exp \left\{
		\Im z \cdot \arctan( 2  \mu / u)\right\}
\end{eqnarray*}
is bounded for any \(u \geq 0 \) iff \(\mu \; \Im(z) <0\), and the function
\begin{eqnarray*}
\left| 
\left[
\overline{
	\psi(u)
}
\right]^{z}
\right| = 
	\exp\left\{
		(\Re(z)/2) \log \left(
			\mu^{2} u^{2} 
			+
			u^{4} / 4
		\right)
	\right\}
	\cdot  
	\exp \left\{
		\Im z \cdot \arctan(- 2  \mu / u)\right\}
\end{eqnarray*}
is bounded for any \(u \geq 0 \)  iff \(\mu \; \Im(z) >0\). Therefore, both integrals in \eqref{hatM} converge. Moreover, note that this estimate possesses the property \(\overline{\M \left[
	\L_{\xi}
\right]  (\overline{z}) }= \M \left[
	\L_{\xi}
\right]  (z),\) which also holds for the original Mellin transform  \(\M \left[
	\L_{\xi}
\right]  (z).\)
\par
The Mellin transform  \(\M \left[
	\L_{\xi}
\right](z)\) is closely connected to the Mellin tranform of the density \(g.\) Indeed,
\begin{eqnarray*}
\M \left[
	\L_{\xi}
\right](z) &=& \int_{\R_{+}} u^{z-1} \L_{\xi}(u)du =\int_{\R_{+}} u^{z-1} \left( 
\int_{\R_{+}}
e^{-su} g(s) ds
\right) 
du\\
&=& 
\int_{\R_{+}} 
g(s)  \left( 
\int_{\R_{+}}
e^{-su}  u^{z-1} du
\right) 
ds\\
&=& 
\Gamma(z) \cdot 
\int_{\R_{+}} 
g(s) s^{-z}
ds=
\Gamma(z) \cdot \M \left[
g
\right] (1-z).
\end{eqnarray*}
Therefore, the Mellin transform of the density of the r.v. \(\xi\) can be represented as 
\begin{eqnarray}
\label{main}
\M \left[
g
\right](z)
=\frac{\M \left[
	\L_{\xi}
\right](1-z)}{\Gamma(1-z)}
=
\frac{1}{\Gamma(1-z)}  
\int_{\R_{+}}   \phi_{X}(u) \left[\psi(u)\right]^{-z} \psi'(u)\, du.
\end{eqnarray}
Using the last expression and taking into account \eqref{hatM}, we  define the estimate of \(\M \left[
	g
\right](z)\) by 
\begin{eqnarray*}
\widehat{\M} \left[
	g
\right](z)  
:=
\frac{\widehat{\M} \left[
	\L_{\xi}
\right](1-z)  
}{\Gamma(1-z)}.
\end{eqnarray*}
Finally, we apply the inverse Mellin transform  to estimate the density \(g\) of the r. v. \(\xi.\) Since the inverse Mellin transform of \(\M \left[
	g
\right](\gamma + \i v) \) is given by 
\begin{eqnarray*}
g(x) = \frac{1}{2 \pi} \int_{\R}
\M \left[
	g
\right](\gamma + \i v)  \cdot   x^{-\gamma-\i v } dv
\end{eqnarray*}
for any \(\gamma \in (0,1),\) we define the estimate of the mixing density \(g\) via 
\begin{eqnarray*}
\widehat{g}_{n,\gamma}^{\circ}(x) &:=& \frac{1}{2 \pi} \int_{-V_{n}}^{V_{n}}
\widehat{\M} \left[
	g
\right](\gamma + \i v)  \cdot   x^{-\gamma-\i v } dv
\\
&=&
\frac{1}{2\pi n}
\sum_{k=1}^{n}
\int_{0}^{V_{n}}
\left[
\int_{0}^{U_{n}}  e^{-\i u X_{k}} \left[ \overline{\psi(u)} \right]^{-\gamma - \i v}  \overline{\psi'(u)} du
\right]
\cdot   \frac{x^{-\gamma-\i v }}{\Gamma(1-\gamma-\i v)} dv\\
&& + \frac{1}{2\pi n}
\sum_{k=1}^{n}
\int_{-V_{n}}^{0}
\left[
\int_{0}^{U_{n}}  e^{\i u X_{k}} \left[\psi(u)\right]^{-\gamma - \i v}  \psi'(u) du
\right]
\cdot   \frac{x^{-\gamma-\i v }}{\Gamma(1-\gamma-\i v)} dv
\end{eqnarray*}
for some \(\gamma\in (0,1)\) and a sequence \(V_n\to\infty\) as \(n\to \infty.\) The convergence rates of the estimate \(\widehat{g}_{n,\gamma}^{\circ} \) crucially depend on the asymptotic behavior of the Mellin transform of the true density function \(g.\) In order to specify this behavior, we introduce two classes of probability densities:
\begin{eqnarray}
\label{E}
\mathcal{E}(\alpha,\gamma_{\circ}, \gamma^{\circ},L) & := & \left\{ p:\,\sup_{\gamma \in (\gamma_{\circ}, \gamma^{\circ})}\int_{\R} e^{\alpha|v|}\left|\mathcal{M}[p](\gamma+\i v)\right| dv \leq L\right\},\\
\label{P}
\mathcal{P}(\beta,\gamma_{\circ}, \gamma^{\circ},L) & := & \left\{ p:\,\sup_{\gamma \in (\gamma_{\circ}, \gamma^{\circ})}\int_{\R} |v|^{\beta}\left|\mathcal{M}[p](\gamma+\i v)\right| dv \leq L\right\},
\end{eqnarray}
where \(\alpha, \beta \in \R_{+},\)  \(L>0\), \(0<\gamma_{\circ}<\gamma^{\circ}<1\). For instance, the gamma-distribution belongs to the first class, and the beta-distribution - to the second, see \cite{panovbel2015}.

The following convergence rates are proved in Section~\ref{proof:convg}.
\begin{thm}
\label{thm41}
Let \(U_{n}=n^{1/4}\) and \(V_{n} = \kappa \ln(n)\) for some \(\kappa>0.\) 
\begin{enumerate}[(i)]
\item If \(g \in \mathcal{E}(\alpha,\gamma_{\circ}, \gamma^{\circ},L) \) for some \(\alpha \in \R_{+}, L>0\), \(0<\gamma_{\circ}<\gamma^{\circ}<1/2,\) then under the choice
\(\kappa=\gamma^{\circ}/(\pi+2\alpha),\) it holds
 for any \(x \in \R_{+},\)
\begin{eqnarray*}
\sqrt{
\E\left[(|x|^{2\gamma_{\circ}}\wedge 1) \cdot
\left|
	 \widehat{g}_{n,\gamma}^{\circ}(x) - g(x)
\right|^{2}
\right]} &\lesssim& 
n^{- \alpha \gamma^{\circ} /(\pi + 2 \alpha)}, \quad n\to \infty, 
\end{eqnarray*}
for any \(\gamma\in (\gamma_{\circ},\gamma^{\circ}),\) where $\lesssim$ stands for an inequality up to some positive finite constant
depending on \(\alpha,\gamma_{\circ}, \gamma^{\circ}\) and \(L.\)
\item 
If \(g \in \mathcal{P}(\beta,\gamma_{\circ}, \gamma^{\circ},L) \) for some \(\beta \in \R_{+}, L>0\), \(0<\gamma_{\circ}<\gamma^{\circ}<1/2,\) then  for any \(\kappa>0\)  and any \(x \in \R_{+},\)
\begin{eqnarray*}
\sqrt{
\E\left[(|x|^{2\gamma_{\circ}}\wedge 1)  \cdot
\left|
	 \widehat{g}_{n,\gamma}^{\circ}(x) - g(x)
\right|^{2}
\right]} &\lesssim& 
\log^{-\beta}(n),\quad n\to \infty,
\end{eqnarray*}
for any \(\gamma\in (\gamma_{\circ},\gamma^{\circ}),\) where $\lesssim$ stands for an inequality up to some positive finite constant
depending on \(\beta,\gamma_{\circ}, \gamma^{\circ}\) and \(L.\)
\end{enumerate}
\end{thm}
\section{Estimation of $G$ with unknown $\mu$} 
\label{sec5}
Using the same strategy as in the previous section and substituting  the true value \(\mu\) by the estimate \(\hmu_{n},\) we arrive at the following estimate of the density function \(g\) in the case of an unknown \(\mu:\)
\begin{multline}\label{gngamma}
\widehat{g}_{n,\gamma}(x) 
=\\
\frac{1}{2\pi n}
\sum_{k=1}^{n}
\int_{0}^{V_{n}}
\left[
\int_{0}^{U_{n}}  e^{-\i u X_{k}} \left[ \overline{\hpsi_{n}(u)} \right]^{-\gamma - \i v}  \overline{\hpsi_{n}'(u)} du
\right]
\cdot   \frac{x^{-\gamma-\i v }}{\Gamma(1-\gamma-\i v)} dv\\
 + \frac{1}{2\pi n}
\sum_{k=1}^{n}
\int_{-V_{n}}^{0}
\left[
\int_{0}^{U_{n}}  e^{\i u X_{k}} \left[\hpsi_{n}(u)\right]^{-\gamma - \i v}  \hpsi_{n}'(u) du
\right]
\cdot   \frac{x^{-\gamma-\i v }}{\Gamma(1-\gamma-\i v)} dv,
\end{multline}
where \(\hpsi_{n}(u):= -\i \hmu_{n} u + u^{2}/2, \;u \in \R.\) The next theorem shows that the difference between \(\widehat{g}_{n,\gamma}(x) \) and \(\hg_{n,\gamma}^{\circ} (x)\) is basically of order \(\hmu_{n} - \mu.\)
\begin{thm}
\label{thm511}
Let the assumptions of Theorem~\ref{thm41} be fulfilled, \(U_{n}=n^{1/4}, \; V_{n} = \kappa \ln(n)\) for some \(\kappa>0\) and \(\mu \ne 0\). Furthermore,  let  \(\hmu_{n}\) be a consistent estimate of \(\mu.\) 
Then for any \( x \in \R, \)
\begin{eqnarray}
\label{thm51}
\sqrt{
	\E 
	\left[(|x|^{2\gamma_{\circ}}\wedge 1) \cdot
	\left|
		 \widehat{g}_{n,\gamma}(x) - \hg_{n,\gamma}^{\circ} (x)
	\right|^{2}
	\right]
}
=
		\beta_{n}\cdot
\bigl\|
		\hmu_{n} - \mu
\bigr\|_{2}
	+ 
	R_{n}, 
\end{eqnarray}
for any \(\gamma\in (\gamma_{\circ},\gamma^{\circ}),\) where 
 \begin{eqnarray*}
R_{n} =
	\delta_{n}\cdot
	\left[
\bigl\|
	\hmu_{n} - \mu
\bigr\|_{4}^{2}		
	+
\bigl\|
	\hmu_{n} - \mu
\bigr\|_{6}^{3}		
\right]
\end{eqnarray*}
and \(\beta_{n}, \delta_{n}\) are positive deterministic sequences such that 
\begin{eqnarray*}
\beta_{n} \lesssim n^{-(3/8) + \pi \kappa/2} (\log(n))^{1/2},\quad
\delta_{n} \lesssim  n^{-(3/8) + \pi \kappa} (\log(n)),
\end{eqnarray*}
as \(n\to \infty,\) where $\lesssim$ stands for an inequality with some positive finite constant
depending on the parameters of the corresponding class. In particular, in the setup of Theorem~\ref{thm41}(i),  \(\beta_{n} \lesssim n^{-1/8} (\ln(n))^{1/2}, \) \(\delta_{n} \lesssim n^{1/8}\ln(n).\)
\end{thm}
\begin{cor}
In the setup of Theorem~\ref{thm:murates}, it holds 
\begin{eqnarray*}
	\sqrt{\E 
	\left[(|x|^{2\gamma_{\circ}}\wedge 1)\cdot
	\left|
		 \widehat{g}_{n,\gamma}(x) - \hg_{n,\gamma}^{\circ} (x)
	\right|^{2}
	\right]
	}
	\lesssim n^{-1/2}, \quad n\to \infty. 
\end{eqnarray*}
for any \(\gamma\in (\gamma_{\circ},\gamma^{\circ}).\)
\end{cor}

%

\section{Numerical example}\label{num}
In this section, we illustrate the performance of estimation our algorithm in  the case, when \(G\) is the distribution function of  the so-called generalized inverse Gaussian distribution  \(GIG \left(
\lambda, \delta, \psi
\right)\)  with a density
 \begin{eqnarray*}
g(x) = \frac{\left(
\psi/\delta
\right)^{\lambda}}{2 K_{\lambda}\left(
\delta \psi
\right)} 
x^{\lambda-1}
\exp\left\{
	- \frac{1}{2} \left(
		\psi^{2} x + \delta^{2} x^{-1}
	\right)
\right\}, \qquad x>0,
\end{eqnarray*}
where \(\lambda \in \R, \delta>0, \psi>0, \) and \(K_{\lambda}(x) = (1/2) \int_{\R_{+}} u^{\lambda-1} \exp \left\{
	u+u^{-1}
\right\} du\) is the Bessel function of the third kind. Trivially,  \(GIG \left(
\lambda, \delta, \psi
\right)\) is an exponential class of distributions. Furthermore, it is interesting to note that 
these distributions are  self-decomposable, see \cite{Halgreen}, and therefore infinitely divisible. 

\begin{figure}
\begin{center}
\includegraphics[width=1\linewidth ]{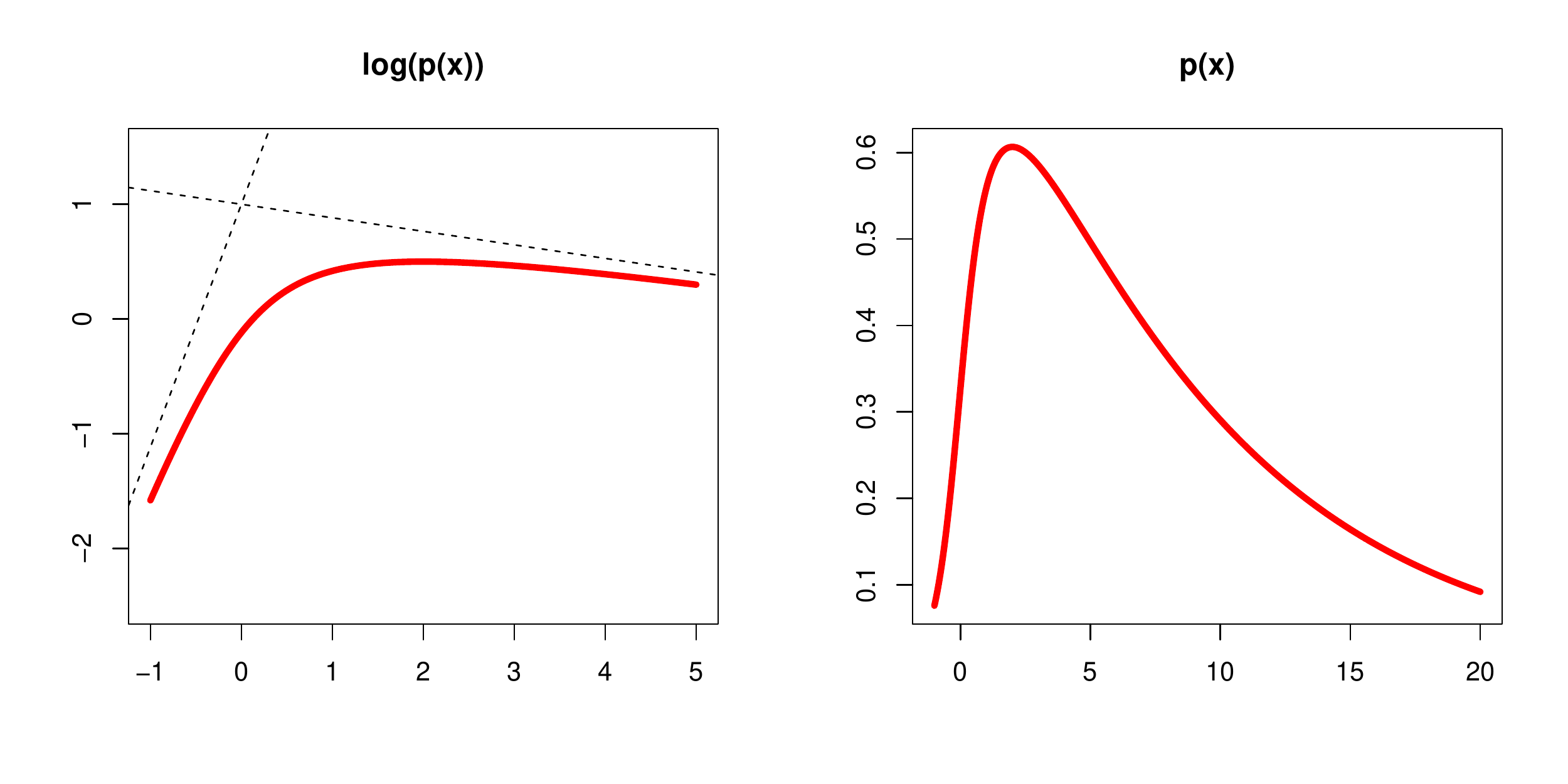}\caption{\label{fig1} Plots of log-density and density of the generalized hyperbolic distibution with parameters $(\sqrt{5}/2,1,1,1)$. Two asymptotes on the left plot correspond to the lines $y=1+(1-\sqrt{5}/2)x$ and $y=1+(1+\sqrt{5}/2)x$.}
\end{center}
\end{figure}

\begin{figure}
\begin{center}
\includegraphics[width=0.45\linewidth ]{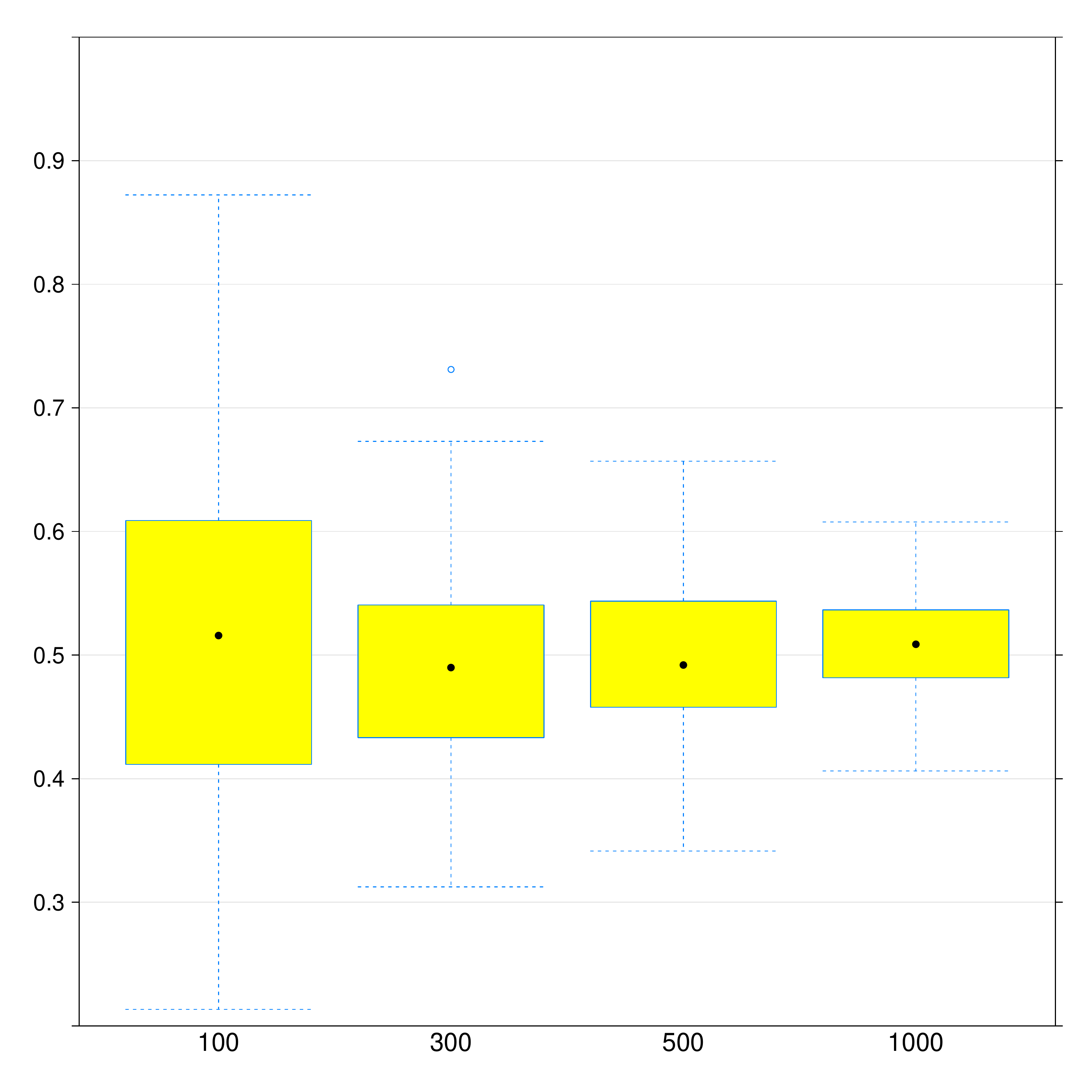}
\includegraphics[width=0.45\linewidth ]{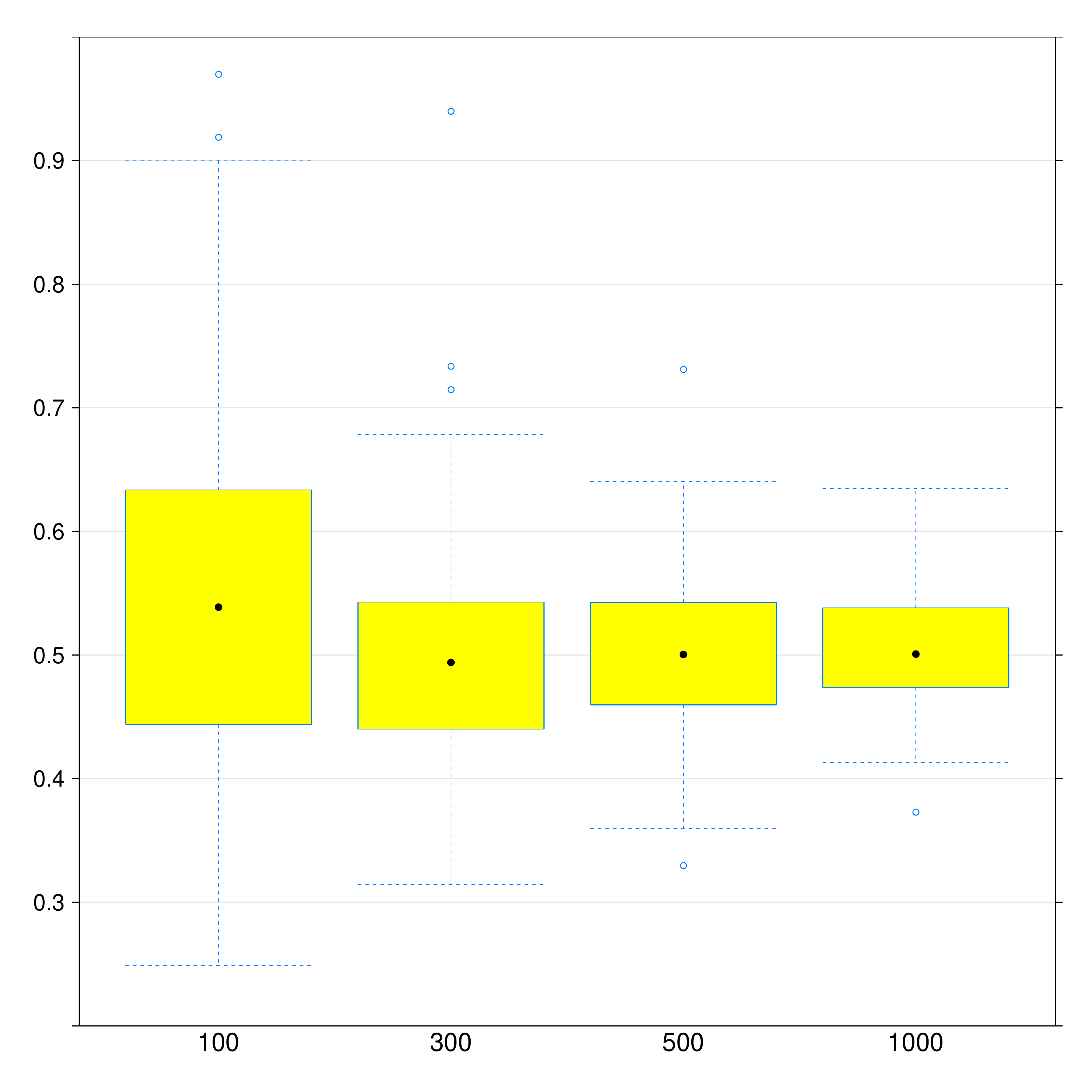}
\caption{\label{fig2} Boxplots of the estimate \( \widehat{\mu}_{n}\) based on \(n=100, 300, 500, 1000\) observations of the r.v. \(X\) under the  inverse gaussian mixing distribution with parameters \(\lambda=\delta=\psi=1\) (left) and \(\lambda=3, \delta=\psi=1\) (right). The true value of the parameter \(\mu\) is \(\mu=0.5\).}
\end{center}
\end{figure}

With this choice of the mixing distribution \(G\), the random variable \(X\) defined by \eqref{X}, has the so-called generalized hyperbolic distribution,  GH \(\left(
\alpha, \lambda, \delta, \psi
\right)\) (\(\alpha = \sqrt{\psi^{2} + \mu^{2}}\)), with  a density function, which can be explicitly computed via \eqref{vm}. In particular, in the case \(\lambda=1,\) the density function is of the form 
\begin{eqnarray*}
p(x)
	&=&
	\frac{\psi} {2 \alpha \delta K_{1}(\delta \psi)}
	\exp \left\{
	- \alpha \sqrt{
		\delta^{2} + 
			x^{2}
	}
	+ \mu
		x 
\right\}.
\end{eqnarray*}
 It would be an interesting to note that the plot of the log-density has two asymptotes \(y=[\log(\psi)-\log(2 \alpha \delta K_{1}(\delta \psi))]+(\mu\pm\alpha)x\), see Figure~\ref{fig1}. For some other properties of this distribution, we refer to \cite{BK}.
The aim  of this simulation study is to estimate \(\mu\) and \(g\) based on the  observations \(X_{1},\ldots, X_{n}\) of the r.v. \(X\). 
Following the idea of Section~\ref{seq:mu}, we first choose the odd weighting function 
\[
w(x) =-\sin(x)\cdot 1(|x|\leq \pi).
\]
Note that \( w(x)\) is bounded and supported on \([-\pi,\pi].\) For our  numerical study, we take \(\lambda=\delta=\psi=1,\) and \(\mu=1/2.\) The boxplots of the estimate \( \widehat{\mu}_{n}\) based on \(100\) simulation runs are presented on Figure~\ref{fig2}. 

\begin{figure}
\begin{center}
\includegraphics[width=0.45\linewidth ]{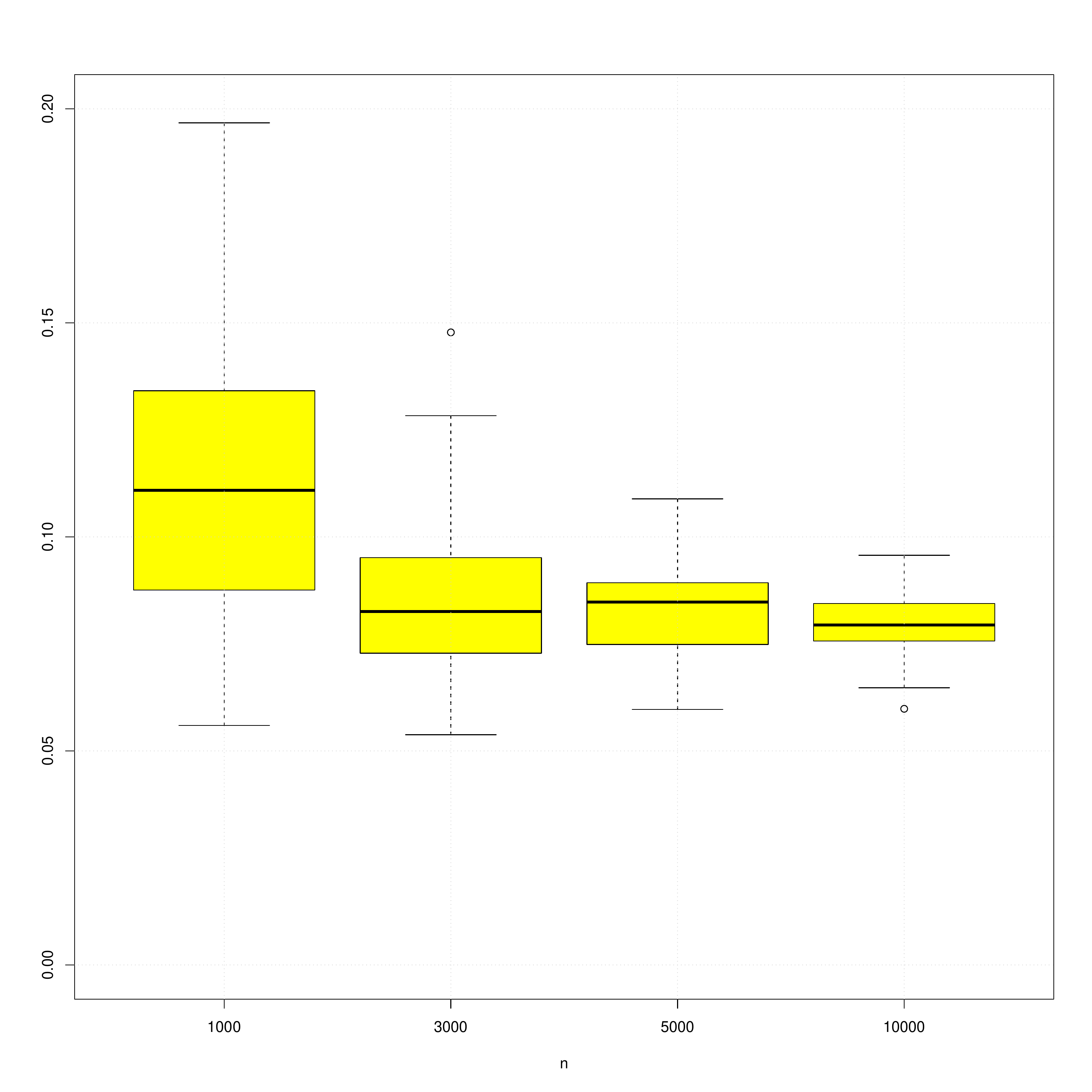}
\includegraphics[width=0.45\linewidth ]{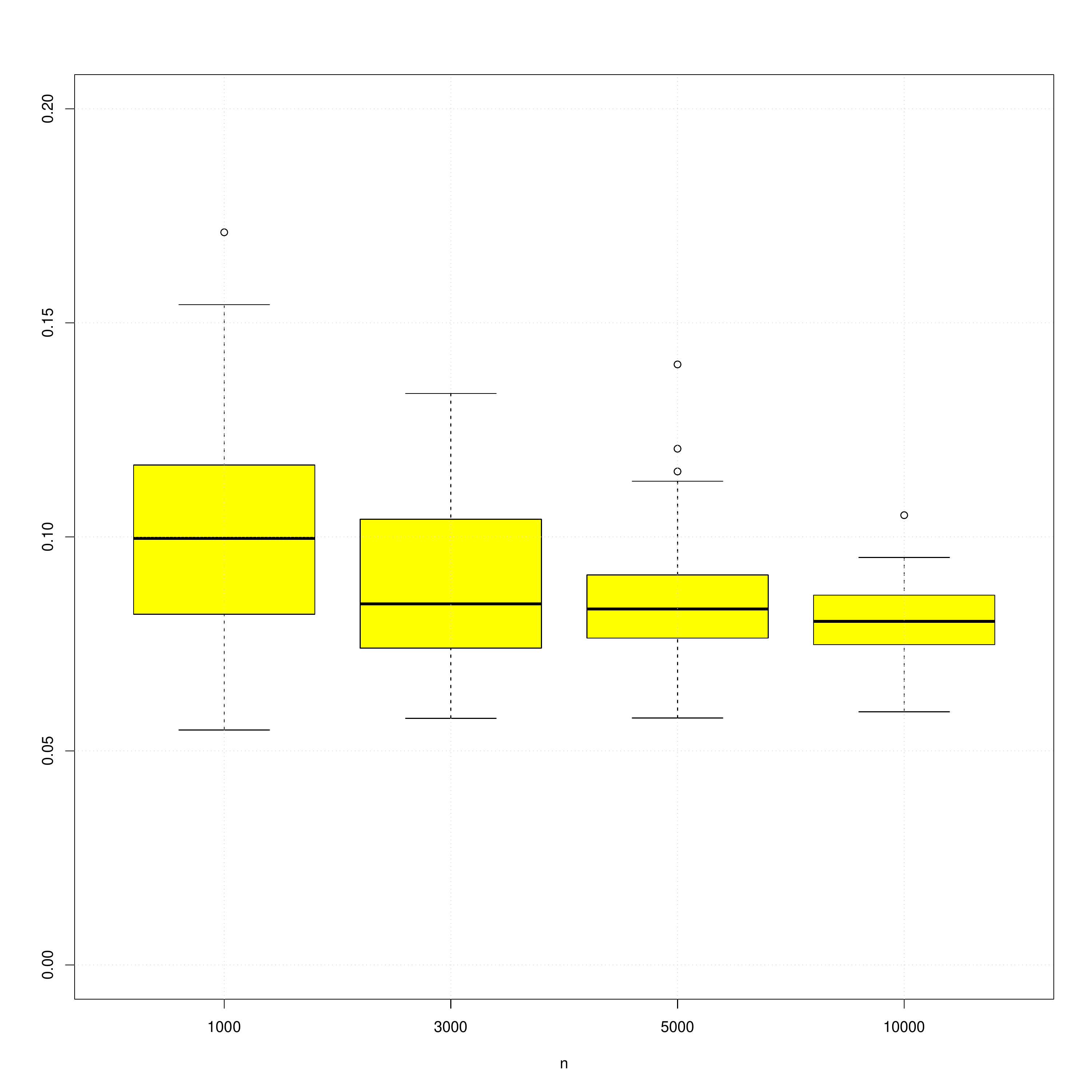}
\caption{\label{fig3} Boxplots of the quantities  \(R( \widehat{g}^{\circ}_n)\) and 
\( R(\widehat{g}_n)\) (corresponding to the cases of known/unknown \(\mu\)) 
based on \(n=1000, 3000, 5000, 10000\) observations of the r.v. \(X\) under the  inverse gaussian mixing distribution with parameters \(\lambda=\delta=\psi=1\).}
\end{center}
\end{figure}

Next, we estimate the density function \(g(x)\) for \(x\in\{x_{1},\ldots,x_{M}\},\) where \(\{x_{1},\ldots,x_{M}\}\) constitute an equidistant grid on \([0.1, 5].\) To this end, we use the estimate  constructed in Section~\ref{sec4}, 
\begin{eqnarray*}
\widehat{g}_{n,\gamma}^{\circ}(x) &=&
\frac{1}{2\pi}
\int_{0}^{V_{n}}
\left[
\int_{0}^{U_{n}}  \widehat\phi_{n}(-u) \left[ \overline{\psi(u)} \right]^{-\gamma - \i v}  \overline{\psi'(u)} du
\right]
\cdot   \frac{x^{-\gamma-\i v }}{\Gamma(1-\gamma-\i v)} dv\\
&& + \frac{1}{2\pi }
\int_{-V_{n}}^{0}
\left[
\int_{0}^{U_{n}}  \widehat\phi_{n}(u) \left[\psi(u)\right]^{-\gamma - t\i v}  \psi'(u) du
\right]
\cdot   \frac{x^{-\gamma-\i v }}{\Gamma(1-\gamma-\i v)} dv,
\end{eqnarray*}
where \(\widehat\phi_{n}(u) = n^{-1} \sum_{k=1}^{n}e^{-\i u X_{k}}\) is the empirical characteristic function of the random variable \(X\).  The error of estimation is measured by 
\[
R(\widehat{g}_n^{\circ}):=\sqrt{\frac{1}{M}\sum_{k=1}^{M} \left(
\widehat{g}_n^{\circ}(x_{k}) 
-
g(x_{k})
\right)^{2}}.
\]
We take \(\gamma=0.1\) and the parameters \(U_{n}\) and \(V_{n}\)   are chosen by numerical optimization of the functional \(R(\widehat{g}_n^{\circ}),\) which yields in our case  the values \(U_{n}=7.6, \) and \(V_{n}=0.9.\) Following the ideas of Section~\ref{sec5}, we consider also the estimate \(\widehat{g}_{n,\gamma}(x), \) which is obtained from \(\widehat{g}_{n}^{\circ}(x)\) by replacing \(\mu\) with its estimate \(\hmu_{n},\) see \eqref{gngamma}.
The difference between \(\widehat{g}^{\circ}_{n,\gamma}(x) \) and \(\widehat{g}_{n,\gamma}(x) \) (which was theoretically considered in Theorem~\ref{thm511}) is illustrated by boxplots on Figure~\ref{fig3}, which shows that the quality of these estimates is essentially the same.
\section{Real data example}
\label{rdex}
In this section, we provide an example of the application of  our model \eqref{vm} for describing   the diamond sizes in marine deposits in South West Africa. The motivation for using this model in this problem can be found in the  paper by Sichel \cite{sichel}: \textit{``According to one geological theory diamonds were transported from inland down the Orange River... One would expect then that the diamond traps would catch the larger stones preferentially and that the average stone weights would decrease as distance from the river mouth increased. Intensive sampling has actually proved this hypothesis to be correct...''} 

Later, Sichel claims that although  \textit{``for relatively small mining areas, and particular for a single-trench unit, the size distributions appear to follow the two-parameter lognormal law,''}  for large mining areas,  one should expect that the parameters of the lognormal law depend on the distance from the mouth of the river. Moreover, taking into account the geological studies, it is reasonable to assume that these parameters  are related inversely to the distance from the mouth of the river and related directly to each other.  Based on these ideas, Sichel proposes to use the model  \eqref{vm} (or, more precisely, a slightly more general model \eqref{Kors}) with \(G\) corresponding to the gamma distribution. Later, 
Barndorff-Nielsen \cite{BN77} applied the same model with \(G\) corresponding to the generalized inverse Gaussian distribution, which was presented above in Section~\ref{num}.

\begin{figure}
\begin{center}
\includegraphics[width=1\linewidth ]{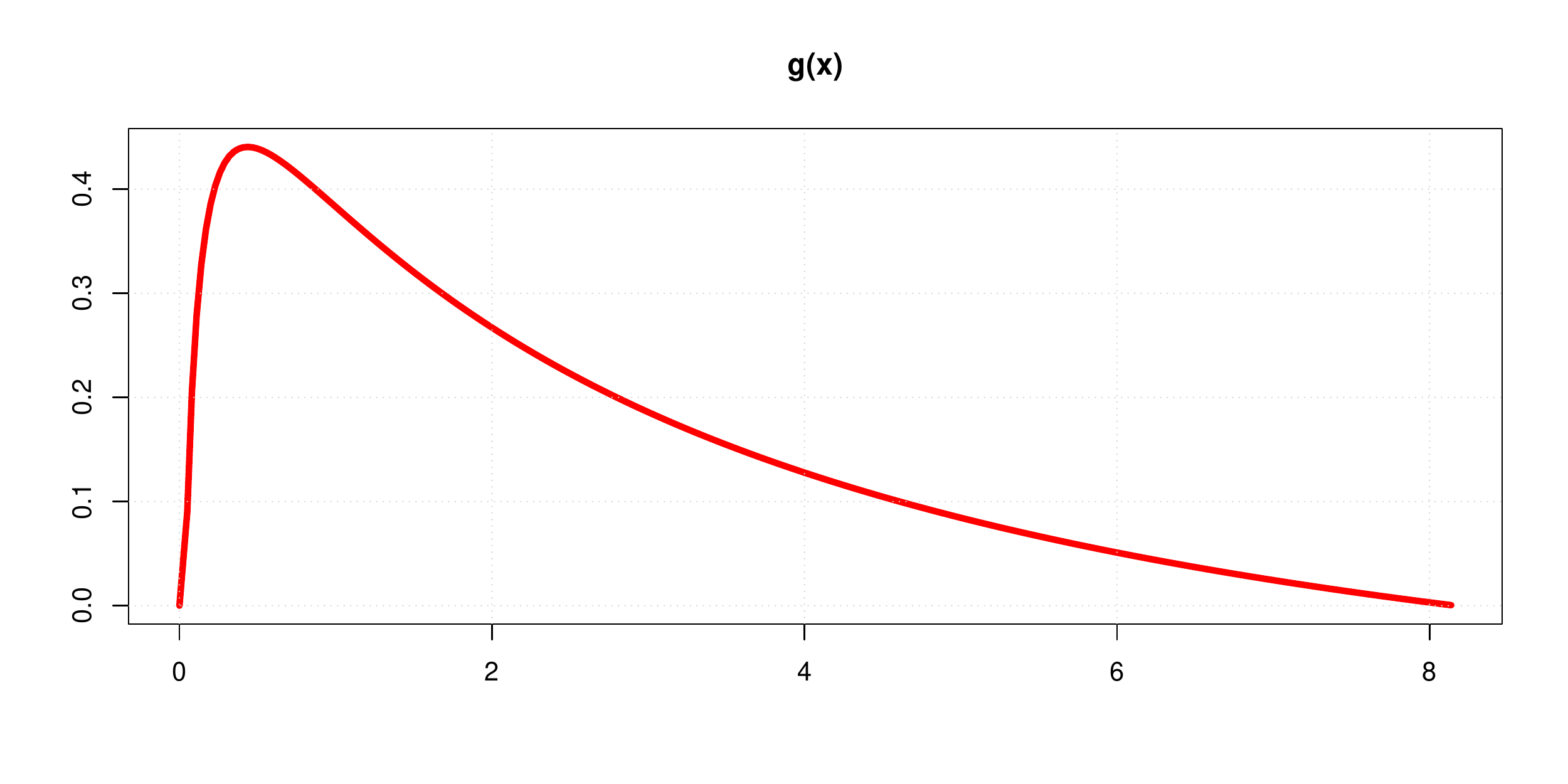}\caption{
	\label{picg}Plot of the  estimate \(\widehat{g}_{n,\gamma}(x)\).
}
\end{center}
\end{figure}

Below we apply our approach to the same data, which can be found (in aggregated form) both in  \cite{BN77} (p. 409) and \cite{sichel} (p. 242). We have 1022 observations of  stone sizes, measured in carats, and aim to fit the model \eqref{vm} to the density of the logarithms of these sizes. The estimation scheme  consists of 2 steps.
\begin{enumerate}
\item First, we estimate the parameter \(\mu\) by  \(\widehat\mu_{n}\) defined in \eqref{wmu}. In this example, we got an estimated value of the parameter  \(\mu\) equal to \(\hat{\mu}_{n}=0.068\). Note that the positive sign of this estimate is important due to the demand on direct relation between the parameters.
\item Second, we estimate the density \(g(s)\) by \(\widehat{g}_{n}(s)\) defined in \eqref{gngamma} for \(s=\{s_{1},\ldots,s_{m}\}\) from the equidistant grid on  \([0.1, 8]\) with step \(\Delta_{s}\). The plot of this function is given as Figure~\ref{picg}.
\end{enumerate}

\begin{figure}
\begin{center}
\includegraphics[width=1\linewidth ]{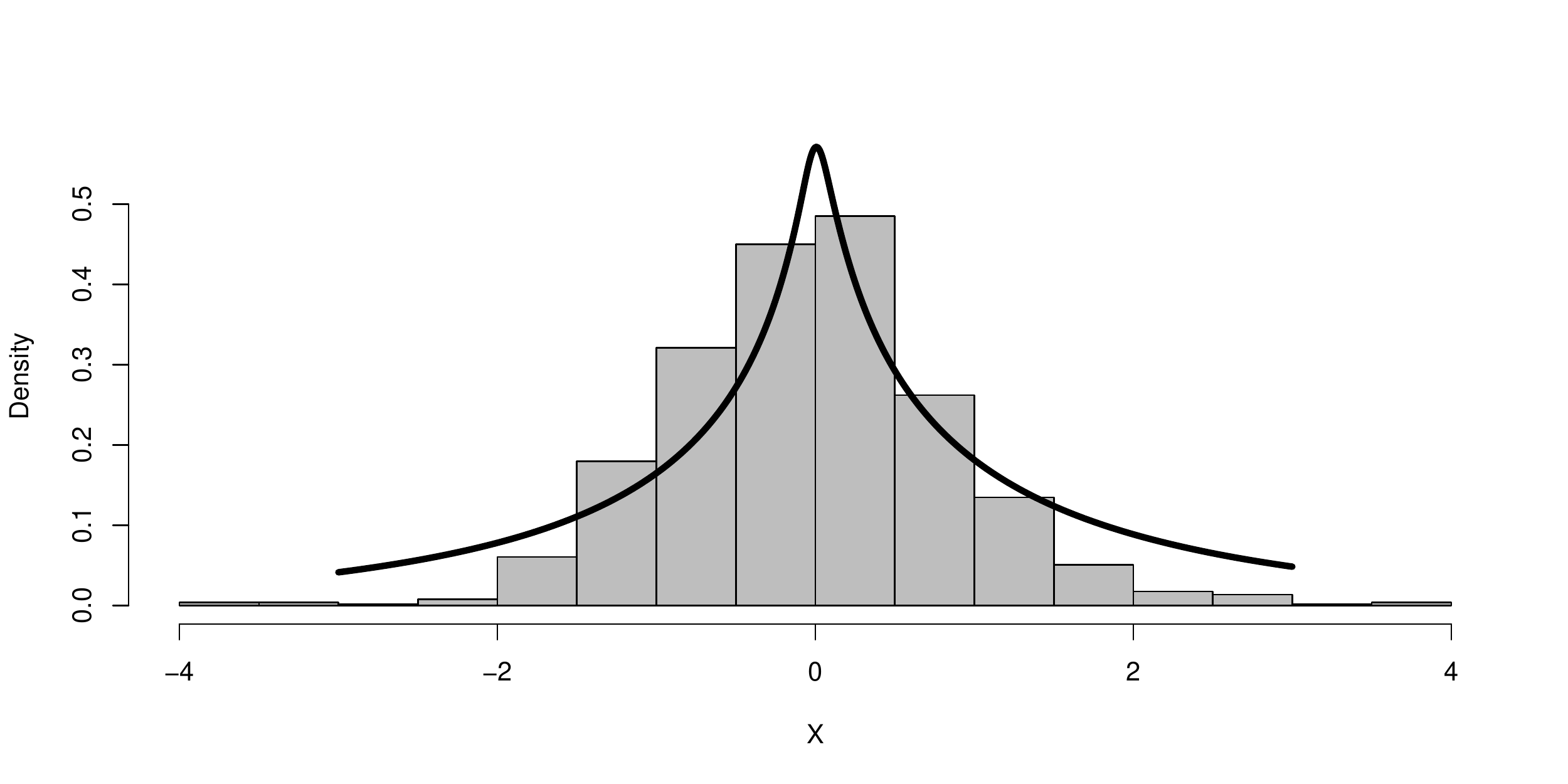}\caption{\label{fig4}Histogram of the logarithms of stone sizes and  the estimate of the density function fitted by the variance-mean mixture model.
}
\end{center}
\end{figure}

To illustrate the performance of our procedure, we also estimate the density fitted by the model \eqref{vm}:
\begin{eqnarray*}
\hat{p}^{\circ}_{n}(x):=
\frac{ \Delta_{s}}{m}
\sum_{k=1}^{m}
\varphi_{\NN(\widehat\mu_{n} s_{k}, s_{k})} (x) \hat{g}_{n}(s_{k}).
\end{eqnarray*}
The performance of this estimate can be visually checked by Figure~\ref{fig4}.

%
%
%

\section{Proofs}
\subsection{Proof of Theorem~\ref{thm:murates}}
Note that 
\begin{eqnarray}\label{m}
\left\Vert \mu_{n}-\mu\right\Vert _{p}\leq
\left\Vert \left(
\mu_{n}-\mu
\right) \I\left\{
\mu_{n} <M
\right\}\right\Vert _{p}
+
\left\Vert \left(
\mu_{n}-\mu
\right) \I\left\{
\mu_{n} =M
\right\}\right\Vert _{p}.\end{eqnarray}
The first summand in the r.h.s. can be bounded by taking 
into account that 
\[
W(\mu_{n})-W(\mu)=W'(\widetilde{\mu}_{n})(\mu_{n}-\mu)
\]
for some $\widetilde{\mu}_{n}\in(\mu\wedge\mu_{n},\mu\vee\mu_{n})$
and hence on the event \(\{ \mu_{n} < M\}\) it holds
\[
\mu_{n}-\mu=\frac{W(\mu_{n})-W_{n}(\mu_{n})}{W'(\widetilde{\mu}_{n})}.
\]
 Note that the function 
\[
W'(\rho)=\mathbb{E}\left[(-X)e^{-\rho X}w(X)\right]
\]
is positive on $\mathbb{R}_{+}$ and attains its minimum at $\rho=\mu$
with 
\[
W'(\mu)=\mathbb{E}\left[(-X)e^{-\mu X}w(X)\right]>0.
\]
 Hence 
 \begin{eqnarray*}
\left\Vert \left(
\mu_{n}-\mu
\right) \I\left\{
\mu_{n} <M
\right\}\right\Vert _{p}
\leq \frac{1}{W'(\mu)}\left\Vert \sup_{\rho\in[-M,M]}\left|W(\rho)-W_{n}(\rho)\right|\right\Vert _{p},
\end{eqnarray*}
and continuing line of reasoning in \eqref{m}, we get
\[
\left\Vert \mu_{n}-\mu\right\Vert _{p}\leq\frac{1}{W'(\mu)}\left\Vert \sup_{\rho\in[-M,M]}\left|W(\rho)-W_{n}(\rho)\right|\right\Vert _{p}+M\left[\mathbb{P}\left(\mu_{n}=M\right)\right]^{1/p}.
\]
Furthermore, due to the monotonicity of $W$ and the fact that $W_{n}(M/2)<0$
on the event $\{\mu_{n}=M\},$ we get 
\[
\mathbb{P}\left(\mu_{n}=M\right)\leq\mathbb{P}\left(W(M/2)-W_{n}(M/2)>W(M/2)>0\right).
\]
Set $\Delta_{n}(\rho):=\sqrt{n}\left(W_{n}(\rho)-W(\rho)\right),$
then
\begin{eqnarray*}
\Delta_{n}(\rho)-\Delta_{n}(\rho') & =& \frac{1}{\sqrt{n}}\sum_{i=1}^{n}\left\{ \left(e^{-\rho X_{i}}-e^{-\rho'X_{i}}\right)w(X_{i})\right. \\
&& \left. \hspace{3cm} -\mathbb{E}\left[\left(e^{-\rho X_{i}}-e^{-\rho'X_{i}}\right)w(X)\right]\right\} 
\end{eqnarray*}
with \(\rho, \rho' \in [0,M].\)
By the Rosenthal and Lyapunov inequalities
\begin{eqnarray*}
\left\Vert \Delta_{n}(\rho)-\Delta_{n}(\rho')\right\Vert _{p} &\leq& \left\Vert \left(e^{-\rho X}-e^{-\rho'X}\right)w(X)\right\Vert _{2}\\
&& \hspace{1cm}+2n^{1/p-1/2}\left\Vert \left(e^{-\rho X}-e^{-\rho'X}\right)w(X)\right\Vert _{p}\\
 &\leq& 3\left|\rho-\rho'\right|\left\Vert \left(e^{-\rho X}+e^{-\rho'X}\right)X w(X)\right\Vert _{p}\\
 &\leq&6\left|\rho-\rho'\right|\left\Vert \left(
1 + e^{-MX}
\right) X w(X)\right\Vert _{p},
\end{eqnarray*}
where the second inequality follows from \(|e^{x} - e^{y}| \leq |x-y| \left(
e^{x} +e^{y}
\right), \forall x,y>0.\)
Next, applying the maximal inequality (see Theorem~8.4 in \cite{Kosorok}), we get 
\[
\left\Vert \sup_{\rho\in[0,M]}\left|\Delta_{n}(\rho)\right|\right\Vert _{p}\leq K'\int_{0}^{M}\left(\frac{1}{\epsilon}\right)^{1/p}\,d\epsilon=KM^{1-1/p}
\]
 for some constants $K$ and $K'$ depending on $p$ and $\left\Vert e^{-MX}w(X)\right\Vert _{p}.$
Hence 
\[
\mathbb{P}\left(W(M/2)-W_{n}(M/2)\geq W(M/2)\right)\leq\frac{K^{p}M^{p-1}}{W^{p}(M/2)\cdot n^{p/2}}.
\]
Finally, we get the result
\begin{align*}
\left\Vert \mu-\mu_{n}\right\Vert _{p} & \leq\frac{1}{W'(\mu)}\left\Vert \sup_{\rho\in[0,M]}\left|W(\rho)-W_{n}(\rho)\right|\right\Vert _{p}+\frac{KM^{1-1/p}}{W(M/2)\cdot n^{1/2}}\\
 & \leq\frac{KM^{1-1/p}}{n^{1/2}}\left[\frac{1}{W'(\mu)}+\frac{1}{W(M/2)}\right].
\end{align*}

\subsection{Proof of Theorem~\ref{thm41}}
\label{proof:convg}
\textbf{1.} The bias of \(\widehat{g}_{n,\gamma}^{\circ}(x)\) 
\begin{multline}
\left|
	\E [\widehat{g}_{n,\gamma}^{\circ}(x)] - g(x)
\right| 
\\=
\left| 
\frac{1}{2\pi}
\int_{0}^{V_{n}}
\left[
\int_{0}^{U_{n}}  \phi_{X}(-u) \left[\overline{\psi(u)}\right]^{-\gamma - \i v}  \overline{\psi'(u)} du
\right]
\cdot   \frac{x^{-\gamma-\i v }}{\Gamma(1-\gamma-\i v)} dv\right. \\
+\frac{1}{2\pi}
\int_{-V_{n}}^{0}
\left[
\int_{0}^{U_{n}}  \phi_{X}(u) \left[\psi(u)\right]^{-\gamma-\i v}\psi'(u) du
\right]
\cdot   \frac{x^{-\gamma-\i v }}{\Gamma(1-\gamma-\i v)} dv\\
\label{bias}
\left.
-
\frac{1}{2 \pi} \int_{\R}
\M \left[
	g
\right](\gamma + \i v)  \cdot   x^{-\gamma-\i v } dv\right|.
\end{multline}
Taking into account \eqref{main}, we get that the last term in this representation can be written as 
\begin{multline*}
 \int_{\R}
\M \left[
	g
\right](\gamma + \i v)  \cdot   x^{-\gamma-\i v } dv 
= \int_{|v|>V_{n}}
\M \left[
	g
\right](\gamma + \i v)  \cdot   x^{-\gamma-\i v } dv  + R_{n},
\end{multline*}
where
\begin{eqnarray*}
R_{n}
&=&
 \int_{|v| \leq V_{n}} 
\left[
	\int_{\R_{+}}   \phi_{X}(u) \left[\psi(u)\right]^{-\gamma - \i v} \psi'(u) du 
\right] \frac{ x^{-\gamma-\i v }}{\Gamma(1-\gamma - \i v)}  dv\\
&=&\int_{0}^{V_{n}}
\left[
	\int_{\R_{+}}   \phi_{X}(-u) \left[\psi(-u)\right]^{-\gamma - \i v} \psi'(-u) du 
\right] \frac{ x^{-\gamma-\i v }}{\Gamma(1-\gamma - \i v)}  dv
\\
&&  +  \int_{-V_{n}}^{0}
\left[
	\int_{\R_{+}}   \phi_{X}(u) \left[\psi(u)\right]^{-\gamma - \i v} \psi'(u) du 
\right] \frac{ x^{-\gamma-\i v }}{\Gamma(1-\gamma - \i v)}  dv.
\end{eqnarray*}
Substituting these expressions into \eqref{bias} and taking into account that \(\overline{\psi(u)} = \psi(-u)\) and \(\overline{\psi'(u)} = -\psi'(-u)\), we derive 
\begin{eqnarray*}
\left|
	\E [\widehat{g}_{n,\gamma}^{\circ}(x)] - g(x)
\right| 
\leq J_{1} + J_{2} +J_{3},
\end{eqnarray*}
where
\begin{eqnarray*}
J_{1}
&:=&
\frac{1}{2\pi}
\left| 
\int_{0}^{V_{n}}
\left[
\int_{U_{n}}^{\infty}  \phi_{X}(-u) \left[\overline{\psi(u)}\right]^{-\gamma - \i v}  \overline{\psi'(u)} du
\right]
\cdot   \frac{x^{-\gamma-\i v }}{\Gamma(1-\gamma-\i v)} dv\right|, \\
J_{2} &:=&
\frac{1}{2\pi}
\left| 
\int_{-V_{n}}^{0}
\left[
\int_{U_{n}}^{\infty}  \phi_{X}(u) \left[\psi(u)\right]^{-\gamma-\i v}\psi'(u) du
\right]
\cdot   \frac{x^{-\gamma-\i v }}{\Gamma(1-\gamma-\i v)} dv\right|,\\
J_{3} &:=& 
\frac{1}{2 \pi}
\left|
 \int_{|v|>V_{n}}
\M \left[
	g
\right](\gamma + \i v)  \cdot   x^{-\gamma-\i v } dv\right|.
\end{eqnarray*}
Upper bound for  \(J_{3}\) directly follows from our assumption on the asymptotic Mellin transform 
\(
\M \left[
	g
\right].\) In the exponential case,
\begin{eqnarray*}
J_{3} 
	&\leq& 
\frac{|x|^{-\gamma}}{2 \pi}
e^{-\alpha V_{n}}
 \int_{|v|>V_{n}}
 \left|
\M \left[
	g
\right](\gamma + \i v)  
\right|
e^{\alpha |v|}
dv
\leq 
\frac{L}{2 \pi} |x|^{-\gamma}
e^{-\alpha V_{n}},
\end{eqnarray*}
whereas in the polynomial case \(J_{3} \leq L (2 \pi)^{-1} |x|^{-\gamma} V_{n}^{-\beta}.\)
To show the asymptotical behavior of  \(J_{1}\), \(J_{2}\), we first derive the upper bound for the characteristic function \(\phi_{X}(u):\)
\begin{eqnarray*}
\left| 
	\phi_{X}(u)
\right| 
=
\left| 
\L_{\xi} \left(
\psi(u)
\right)
\right| 
\lesssim \frac{1}{u^{2}},
\end{eqnarray*}
where the last asymptotic inequality follows from the integration by parts:
\begin{eqnarray*}
\left| 
\L_{\xi} \left(
w
\right)
\right| 
=
\left|
	 \int_{\R_{+}} g(\lambda) e^{-\lambda w} d\lambda
\right|
=
\left|
	\frac{1}{w} \int_{\R_{+}} g(\lambda) d (e^{-\lambda w})
\right| 
\lesssim \frac{1}{|w|}
\end{eqnarray*}
for all \(w \in \C.\) Second, it holds 
\begin{eqnarray*}
\left|\Gamma(w) \right| \geq 
C \left|
	\Im(w)
\right|^{\Re(w)-1/2} 
e^{-|\Im(w)| \pi /2}
\end{eqnarray*}
for any \(w \in \C\) such that \(\Re(w)\geq -2, |\Im(w)| \geq 1\) and some constant \(C>0\), see \cite{AAR}. Therefore,  we get 
\begin{eqnarray*}
J_{1} &\leq& \frac{|x|^{-\gamma}}{2 \pi}\int_{0}^{V_{n}}
\left[
\int_{U_{n}}^{\infty}
\left| 
	\phi_{X}(-u) 
\right|
\left| 
	\psi(u)
\right|^{-\gamma} 
\left|  
	\psi'(u)
\right| 
du
\right]
\cdot   
\frac{1}{\left|\Gamma(1-\gamma-\i v)\right|} dv \\
&\lesssim& 
|x|^{-\gamma} \int_{0}^{V_{n}}
\left[
\int_{U_{n}}^{\infty}
u^{-2 \gamma - 1} du
\right]
\cdot   
\frac{1}{\left| \Gamma(1-\gamma-\i v) \right| } dv \\
&\lesssim& 
|x|^{-\gamma}
U_{n}^{-2 \gamma} \max\left\{V_{n}^{\gamma-1/2},1 \right\} e^{V_{n} \pi /2}
=
|x|^{-\gamma}
U_{n}^{-2 \gamma} e^{V_{n} \pi /2}
.
\end{eqnarray*}
\textbf{2.} Next, we consider the variance of the estimate \(\widehat{g}_{n,\gamma}^{\circ}:\)
\begin{multline*}
\Var(\widehat{g}_{n,\gamma}^{\circ}(x))  \\ 
=
\frac{1}{(2\pi)^{2} n}
\Var\left\{
\int_{0}^{V_{n}}
\left[
\int_{0}^{U_{n}} e^{-\i u X} \left[ \overline{\psi(u)} \right]^{-\gamma - \i v}  \overline{\psi'(u)} du
\right]
\cdot   \frac{x^{-\gamma-\i v }}{\Gamma(1-\gamma-\i v)} dv
\right.\\
\left.+\int_{-V_{n}}^{0}
\left[
\int_{0}^{U_{n}} e^{\i u X} \left[\psi(u)\right]^{-\gamma - \i v}  \psi'(u) du
\right]
\cdot   \frac{x^{-\gamma-\i v }}{\Gamma(1-\gamma-\i v)} dv
\right\}\\
\leq
\frac{|x|^{-2\gamma}}{(2\pi)^{2} n}
\left\{
\int_{-V_{n}}^{V_{n}}
\left[
\int_{0}^{U_{n}}
\sqrt{
\Var\left\{
	G(X, u,v)
\right\}}
du
\right]
\cdot   \frac{1}{\left| \Gamma(1-\gamma-\i v) \right|} dv
\right\}^{2},
\end{multline*}
where 
\begin{eqnarray*}
G(X,u,v) = \begin{cases}
e^{-\i u X} \left[ \overline{\psi(u)} \right]^{-\gamma - \i v}  \overline{\psi'(u)}, &\text{if $v>0$,}\\
e^{\i u X} \left[ \psi(u) \right]^{-\gamma - \i v}  \psi'(u), &\text{if $v<0$.}
\end{cases}\end{eqnarray*}
The last inequality follows from the fact that  for any integrable function 
\(f:\C^{1+m} \to \C\) and any random variable \(Z\) it holds
\begin{eqnarray*}
\Var\left(
	\int f(Z,x) dx
\right)
&=&
\int	\int \cov\left(  f(Z,x), f(Z,y) \right) dx dy\\
&\leq&
\int	\int \sqrt{\Var f(Z,x)} \sqrt{\Var f(Z,y)} dx dy\\
&=&
\left( 
	\int \sqrt{\Var f(Z,x)} dx
\right)^{2}.
\end{eqnarray*}
It would be a worth mentioning that 
\begin{eqnarray*}
\sqrt{
\Var\left\{ 
 e^{\i u X} \left[\psi(u)\right]^{-\gamma - \i v}  \psi'(u) 
 \right\} 
 }
&\leq&
\sqrt{1 - |\phi_{X}(u)|^{2}}
  \left| \psi(u)\right|^{-\gamma }  
  \left| \psi'(u) \right|\\
  &\leq&
  \left| \psi(u)\right|^{-\gamma }  
  \left| \psi'(u) \right|,
\end{eqnarray*}
and moreover it holds 
\begin{eqnarray*}
\int_{0}^{U_{n}}\left| \psi(u)\right|^{-\gamma }  
  \left| \psi'(u) \right| du &=&
\int_{0}^{U_{n}} u^{-\gamma} \left(
	u^{2}/4 + \mu^{2}
 \right)^{-\gamma/2}
 \left(
u^{2} + \mu^{2}
\right)^{1/2}\\
&\leq&
2^{\gamma}
 \left(
	\int_{0}^{\mu}
	+
	\int_{\mu}^{U_{n}}
\right) u^{- 2 \gamma} 
 \left(
u^{2} + \mu^{2}
\right)^{1/2} du\\
&\leq& 2^{\gamma+1/2}
 \left(
 \mu^{2 - 2\gamma}
	+
	\int_{\mu}^{U_{n}} u^{-2 \gamma +1} du
\right) \\
&\lesssim& U_{n}^{2(1-\gamma)},
\end{eqnarray*}
where we use the assumption \(\gamma<1/2\). Finally, we conclude that 
\begin{eqnarray*}
\Var(\widehat{g}_{n,\gamma}^{\circ}(x)) &\leq& 
\frac{|x|^{-2\gamma}}{(2\pi)^{2} n}
\left[
\int_{0}^{U_{n}}
  \left| \psi(u)\right|^{-\gamma }  
  \left| \psi'(u) \right|
du 
\right]^{2} 
\\
&& \hspace{3cm}
\cdot
\left[
\int_{-V_{n}}^{V_{n}}
  \frac{1}{\left| \Gamma(1-\gamma-\i v) \right|} dv
\right]^{2}\\
&\lesssim&
\frac{|x|^{-2\gamma}}{n}
U_{n}^{4(1-\gamma)} e^{V_{n}\pi}
,
\end{eqnarray*}
where we use  that 
\begin{eqnarray}
\label{gammaf}
\int_{-V_{n}}^{V_{n}}
\cdot   \frac{1}{\left| \Gamma(1-\gamma-\i v) \right|} dv
\lesssim
\max\left\{V_{n}^{\gamma-1/2},1 \right\} \cdot e^{V_{n} \pi /2}
=
e^{V_{n} \pi /2}.
\end{eqnarray}
\textbf{3.} Set \(\rho(x)=|x|^{2\gamma},\) then we have the following the bias-variance decomposition: 
\begin{eqnarray*}
\sqrt{
\rho(x) \cdot
\E\left[
\left|
	 \widehat{g}_{n,\gamma}^{\circ}(x) - g(x)
\right|^{2}
\right]} &\leq& 
\sqrt{
\rho(x) \cdot 
\left(
\left|
	\E \widehat{g}_{n,\gamma}^{\circ}(x) - g(x)
\right|^{2}
+
\Var(\widehat{g}_{n,\gamma}^{\circ}(x))
\right)}
\\
&\leq& 
\sqrt{\rho(x)} \cdot
\left|
	\E \widehat{g}_{n,\gamma}^{\circ}(x) - g(x)
\right|
\\
&& \hspace{3cm}+
\sqrt{
\rho(x) \cdot 
\Var(\widehat{g}_{n,\gamma}^{\circ}(x))
}.
\end{eqnarray*}
For instance, in the exponential case, this decomposition yields
\begin{eqnarray*}
\sqrt{
\rho(x) \cdot
\E\left[
\left|
	 \widehat{g}_{n,\gamma}^{\circ}(x) - g(x)
\right|^{2}
\right]} 
&\lesssim& 
\left(
	\sqrt{\rho(x)}|x|^{-\gamma} 
\right)
\cdot
\left(
U_{n}^{-2 \gamma} e^{V_{n} \pi /2}
+
e^{-\alpha V_{n}}
\right.\\
&& \hspace{2cm}
\left.+
n^{-1/2}
U_{n}^{2(1-\gamma)} e^{V_{n}\pi/2}
\right)\\
&\leq&
U_{n}^{-2 \gamma} 
\left(
	1 + n^{-1/2}
U_{n}^{2} 
\right)
e^{V_{n}\pi/2} 
+ 
e^{-\alpha V_{n}}.
\end{eqnarray*}
The last expression suggests the choice \(U_{n}=n^{1/4},\) under which 
\begin{eqnarray*}
\sqrt{
\rho(x) \cdot
\E\left[
\left|
	 \widehat{g}_{n,\gamma}^{\circ}(x) - g(x)
\right|^{2}
\right]} &\lesssim& 
n^{-\gamma/2}e^{V_{n}\pi/2} + e^{-\alpha V_{n}}.
\end{eqnarray*}
Choosing \(V_{n}\) in the form \(V_{n}=\kappa \ln(n),\) we arrive at the desired result. 

\subsection{Proof of Theorem~\ref{thm511}}
It's easy to see that 
\begin{multline}
\label{gg}
	\E 
	\left[
	\left|
		 \widehat{g}_n(x) - \hg_{n}^{\circ} (x)
	\right|^{2}
	\right]
	\\=
	\frac{1}{(2\pi n)^{2}}
	\E 
	\left[
	\left|
\sum_{k=1}^{n} 
\int_{0}^{V_{n}}
\left[
\int_{0}^{U_{n}}  e^{-\i u X_{k}} 
\Lambda^{(1)}_{n}(u,v) du
\right]\cdot   \frac{x^{-\gamma-\i v }}{\Gamma(1-\gamma-\i v)} dv
\right. \right.\\
\left.\left.
+
\sum_{k=1}^{n}
\int_{-V_{n}}^{0}
\left[
\int_{0}^{U_{n}}  e^{\i u X_{k}} 
\Lambda^{(2)}_{n}(u,v) du
\right]\cdot   \frac{x^{-\gamma-\i v }}{\Gamma(1-\gamma-\i v)} dv
	\right|^{2}\right],
\end{multline}
where
\begin{eqnarray*}
\Lambda_{n}^{(1)}(u,v) &:=& \Bigl[ \; \overline{\hpsi_{n}(u)} \; \Bigr]^{-\gamma - \i v} \cdot\overline{\hpsi_{n}'(u)}
	-
	\left[ \; \overline{\psi(u)}\; \right]^{-\gamma - \i v} \cdot \overline{\psi'(u)},\\
\Lambda_{n}^{(2)}(u,v) &:=& \left[\hpsi_{n}(u)\right]^{-\gamma - \i v} \cdot \hpsi_{n}'(u)
	-
	\left[\psi(u)\right]^{-\gamma - \i v} \cdot \psi'(u).
\end{eqnarray*}
Below we consider in details \(\Lambda_{n}^{(1)}(u,v);\) the treatment for the second term follows the same lines. 
Denote 
\begin{eqnarray*}
A =A(u) &:=&  \Bigl[ \; \overline{\psi(u)} \; \Bigr]^{-\gamma - \i v} =
\left( 
	\i  u \mu + u^{2} / 2
\right)^{-\gamma- \i v},
\\
A_{n} =A_{n}(u)&:=& \Bigl[ \; \overline{\hpsi_{n}(u)} \; \Bigr]^{-\gamma - \i v} =
\left( 
	\i  u \hmu_{n} + u^{2} / 2
\right)^{-\gamma- \i v},\\
B =B(u) &:=& \overline{\hpsi_{n}'(u)} = \i \mu + u,\qquad \qquad
B_{n} = B_{n}(u) := \overline{\psi'(u)} = \i \hmu_{n} + u.
\end{eqnarray*}
In this notation, 
\begin{eqnarray*}
\Lambda_{n}^{(1)}(u,v) = A_{n}B_{n} - A B 
=
\left(
A_{n}-A
 \right)
 \left(
B_{n}-B
 \right) 
 + 
 \left(
A_{n}-A
 \right) B + 
 \left(
B_{n}-B
 \right) A. 
\end{eqnarray*}
Note that

\begin{eqnarray*}
A_{n}-A&=& 
\left( 
	 \i  u \hmu_{n} + u^{2} / 2
\right)^{-\gamma- \i v} 
-
\left( 
	 \i  u \mu + u^{2} / 2
\right)^{-\gamma- \i v} \\ 
&=& 
A\cdot
\left(
\left( 
1 +   \frac{  \i \left( \hmu_{n} -\mu\right) }{ \i  \mu + u/2}
\right)^{-\gamma-\i v}
 -1
 \right).
\end{eqnarray*}
Next, applying the  Taylor theorem for the function \(g(x)=(1+z x)^{w}: \R \to \C\) in the vicinity of zero with \(z = \i/(\i \mu +u/2), \; w = -\gamma -\i v, \; x= \widehat\mu_{n}-\mu \in \R,\) we get
\begin{eqnarray}
\label{resi}
\left( 
1 +   \frac{  \i \left( \hmu_{n} -\mu\right) }{ \i  \mu + u/2}
\right)^{-\gamma-\i v}
  =
  1 - \frac{  \i \left( \gamma + \i v\right) }{ \i  \mu + u/2} \left(
\widehat\mu_{n} - \mu
\right)
-
r_{n}(u) \left(
	\widehat\mu_{n} - \mu
\right)^{2},
\end{eqnarray}
where 
\begin{eqnarray*}
r_{n}(u) &=& 
\frac{  \left( \gamma + \i v\right)\left( \gamma + \i v + 1\right) }{2
\left(
	\i  \mu + u/2
\right)^{2} }
\left(
1  + \frac{
\i \theta (\widehat\mu_{n} - \mu)
}
{\i \mu + u/2}
\right)^{-\gamma - \i v -2} \\
&=&
\frac{  \left( \gamma + \i v\right)\left( \gamma + \i v + 1\right) }{2
\left(
	\i  \mu + u/2
\right)^{2} }
\left(
 \frac{
\i \mu + u/2
}
{\i \tilde{\mu} + u/2}
\right)^{(\gamma+2) + \i v}, 
\end{eqnarray*}
where \(\theta \in(0,1)\) and \(\tilde{\mu} = \theta \widehat\mu_{n}+(1-\theta) \mu.\)
Note that uniformly on 
\(u \in [0,U_{n}]\) and \(v \in [-V_{n}, V_{n}]\) it holds
\begin{multline*}
|r_{n}(u)| \leq 
\frac{1}{4}
\left(
\gamma^{2} + v^{2}
\right)^{1/2}
\left(
(\gamma+1)^{2} + v^{2}
\right)^{1/2}
\\
\cdot
\frac{
 (\mu^{2}+u^{2}/4)^{\gamma/2}
}
{
(\tilde\mu^{2}+u^{2}/4)^{\gamma/2+1}
}
\cdot 
\frac{
e^{-v \arctan(u/(2\mu))}
}
{e^{-v \arctan(u/(2\tilde\mu))}}
\lesssim V_{n}^{2} e^{\pi V_{n}},
\end{multline*}
and moreover \(|r_{n}(u) (\i \mu +u) | \lesssim V_{n}^{2} e^{\pi V_{n}}.\)
From \eqref{resi} it follows then
\begin{eqnarray*}
A_{n}-A &=& 
-A\cdot 
   \frac{  \i \left(\gamma + \i v \right) }{ \i  \mu + u/2} 
   \left(
	 \hmu_{n} -\mu
\right) 
-
A \cdot r_{n}(u) \left(
	\widehat\mu_{n} - \mu
\right)^{2}, \; n \to \infty,
\end{eqnarray*}
and 
\begin{eqnarray}
\nonumber
\Lambda_{n}^{(1)} (u,v) &=&  \left( \hmu_{n} -\mu\right) 
 \cdot   \i  A \cdot \left( - \left(
	\gamma + \i v 
\right) 
\frac{
 \i \mu +u 
}
{ \i \mu +u/2 } +1 
\right) \\ 
\nonumber
&& +  
\left( \hmu_{n} -\mu\right)^{2}
A \left(
	- r_{n}(u) \left(
		\i \mu +u
	\right) + 
\frac{
	\gamma+\i v
}
{	\i \mu +u/2}
\right)\\
\label{LL}
&&-\left( \hmu_{n} -\mu\right)^{3}  \i A  r_{n}(u).
\end{eqnarray}
In the sequel we assume for simplicity  that on the second stage (estimation of \(G\))  we use another sample, independent of that was used for the estimation  of \(\mu\).  Substituting \eqref{LL} into \eqref{gg}, we get \eqref{thm51} with 
\begin{eqnarray*}
\beta_{n}^{2} |x|^{-\gamma} &=& 
	\frac{1}{(2\pi n)^{2}}
	\E 
	\left[
	\left|
\sum_{k=1}^{n} 
\int_{0}^{V_{n}}
\left[
\int_{0}^{U_{n}}  e^{-\i u X_{k}}  A \cdot \left( - \left(
	\gamma + \i v 
\right) 
\frac{
 \i \mu +u 
}
{ \i \mu +u/2 } +1 
\right) du
\right] \right. \right.\\
&&
\hspace{7cm}
\left.\left.
\cdot   \frac{x^{-\gamma-\i v }}{\Gamma(1-\gamma-\i v)} dv
\right. \right.\\
&&
\left.\left.
+
\sum_{k=1}^{n}
\int_{-V_{n}}^{0}
\left[
\int_{0}^{U_{n}}  e^{\i u X_{k}} 
  \overline{A} \cdot \left( - \left(
	\gamma + \i v 
\right) 
\frac{
 - \i \mu +u 
}
{ - \i \mu +u/2 } +1 
\right) du
\right]\cdot  \right. \right.\\
&&
\hspace{7cm}
\left.\left. \frac{x^{-\gamma-\i v }}{\Gamma(1-\gamma-\i v)} dv
	\right|^{2}\right].
\\
 &\leq& 
	\frac{1}{2\pi^{2} n}
	\E 
	\left[
	\left|
\int_{0}^{V_{n}}
\left[
\int_{0}^{U_{n}}  e^{-\i u X_{1}}  A \cdot \left( - \left(
	\gamma + \i v 
\right) 
\frac{
 \i \mu +u 
}
{ \i \mu +u/2 } +1 
\right) du
\right] \right. \right.\\
&&
\hspace{7cm}
\left.\left.
\cdot   \frac{x^{-\gamma-\i v }}{\Gamma(1-\gamma-\i v)} dv
\right|^{2} \right.\\
&&
\left.
+
\left|
\int_{-V_{n}}^{0}
\left[
\int_{0}^{U_{n}}  e^{\i u X_{1}} 
  \overline{A} \cdot \left( - \left(
	\gamma + \i v 
\right) 
\frac{
 - \i \mu +u 
}
{ - \i \mu +u/2 } +1 
\right) du
\right]\cdot  \right. \right.\\
&&
\hspace{7cm}
\left.\left. \frac{x^{-\gamma-\i v }}{\Gamma(1-\gamma-\i v)} dv
	\right|^{2}\right].
\end{eqnarray*}
Note that due to the Minkowski inequality , 
\begin{eqnarray*}
\beta_{n}^{2} |x|^{-\gamma}&\leq& 
	\frac{1}{2 \pi^{2} n}
	\left(
\int_{-V_{n}}^{V_{n}}
\left[
\int_{0}^{U_{n}}  
	| \phi_{X}(2u) |^{1/2}  | A(u) | \cdot \left| - \left(
	\gamma + \i v 
\right) 
\frac{
 \i \mu +u 
}
{ \i \mu +u/2 } +1 
\right| du
\right]  \right.\\
&&
\hspace{7cm}
\left.
\cdot   \frac{|x|^{-\gamma }}{|\Gamma(1-\gamma-\i v)|} dv
\right)^{2} .
\end{eqnarray*}
Taking into account that \(|\phi_{X}(\cdot)| \leq 1,\) 
\(\left| - \left(
	\gamma + \i v 
\right) 
\left(
 \i \mu +u 
\right)
\left( \i \mu +u/2 \right)^{-1} +1 
\right|  \lesssim 2 V_{n}+1,\) and moreover
\[\int_{0}^{U_{n}}|A(u)| du=\int_{0}^{U_{n}} | u|^{-\gamma} \left(
\mu^{2} +(u^{2}/4)
\right)^{-\gamma/2} e^{v \arctan(u/(2\mu))}du
\lesssim U_{n}^{1-\gamma}e^{ V_{n} \pi/2},\]
we get that 
\begin{eqnarray*}
\beta_{n}^{2}&\lesssim& 
n^{-1} 
U_{n}^{1-\gamma}V_{n} e^{ V_{n} \pi/2}
\cdot
\int_{-V_{n}}^{V_{n}}
\frac{1}{|\Gamma(1-\gamma-\i v)|} dv
\leq n^{-1} 
U_{n}^{1-\gamma} V_{n} e^{ V_{n} \pi},
\end{eqnarray*}
where we use the inequality \eqref{gammaf}. Therefore, under our choice of \(U_{n}\) and \(V_{n}\), we get \(\beta_{n}^{2}\lesssim n^{-(3/4) + \pi \kappa}\ln(n).\)  Analogously, 
\begin{eqnarray*}
\delta_{n}^{2} &\lesssim& n^{-1} 
U_{n}^{1-\gamma} V_{n}^{2} e^{ 2 V_{n} \pi} \lesssim n^{-(3/4) + 2\pi \kappa} (\ln(n))^{2}.
\end{eqnarray*}
This observation completes the proof.

\bibliographystyle{plain}
\bibliography{Panov_bibliography-1}

\end{document}